\documentclass[twocolumn,showpacs,preprintnumbers,amsmath,amssymb,floatfix]{revtex4}

\usepackage{graphicx}
\usepackage{dcolumn}
\usepackage{bm}
\usepackage[usenames]{color}
\usepackage{latexsym}

\makeatletter
\def\lsim{\mathrel{\mathpalette\gls@align<}}
\def\gsim{\mathrel{\mathpalette\gls@align>}}
\def\gls@align#1#2{\lower.6ex\vbox
	{\baselineskip\z@skip\lineskip\z@\ialign
		{$\m@th#1\hfill##$\crcr#2\crcr\sim\crcr}}}
\makeatother

\def\be{\begin{equation}}
\def\ee{\end{equation}}
\def\bea{\begin{eqnarray}}
\def\eea{\end{eqnarray}}
\makeatletter
\def\leqq{\mathrel{\mathpalette\gle@align<}}
\def\geqq{\mathrel{\mathpalette\gle@align>}}
\def\gle@align#1#2{\lower.6ex\vbox{%
\baselineskip\z@skip\lineskip\z@\ialign{$\m@th#1\hfill##$\crcr#2\crcr=\crcr}}}
\makeatother

\begin{document}

\preprint{
Effect of Two-Dimensionality
on Step Bunching Induced by the Drift of Adatoms
}
\title{
Effect of Two-Dimensionality
on Step Bunching Induced by the Drift of Adatoms
}

\author{Masahide Sato$^{1,}$}\email{sato@cs.s.kanazawa-u.ac.jp}
\author{Makio Uwaha$^2$}
\author{Yukio Hirose$^3$}
\affiliation{
    $^1$Information Media Center of Kanazawa University,
		Kakuma-cho, Kanazawa 920-1192, Japan\\
	$^2$Department of Physics, Nagoya University,
		Furo-cho, Chikusa-ku, Nagoya 464-8602, Japan \\
    $^3$Department of Computational Science, Kanazawa University,
		Kakuma-cho, Kanazawa 920-1192, Japan
}

\date{\today}
\begin{abstract}
We study 
the effect of two-dimensionality
on step bunching induced by drift of adatoms.
When anisotropy of the diffusion coefficient changes alternately
on consecutive terraces like a Si(001) vicinal face,
bunching occurs with the drift of adatoms.
If the fluctuation  of step bunches is neglected
as in the one-dimensional model,
the bunching with step-down drift is faster than that with step-up drift
in contradiction with the experiment by Latyshev and coworkers.
In a two-dimensional model, the step bunches 
wander heavily with step-up drift 
and recombination with neighboring bunches occur
more frequently than  those with step-down drift
and the bunching is accelerated.
When the difference of kinetic coefficients between
two types of steps is taken into account,
the bunching with step-up drift can be faster than that with step-down drift.

\end{abstract}


\maketitle

\section{Introduction}
The  Si(001) surface is reconstructed by dimerization of surface atoms.
When its vicinal face is tilted in the $\langle 110 \rangle$ direction,
the terraces  with dimer rows  parallel to the steps (we call $\mathrm{T_A}$)
and those with dimer rows perpendicular to the steps (we call $\mathrm{T_B}$)
appear alternately.
Since the surface diffusion along the dimer rows is faster than
that perpendicular to the dimer rows,
the anisotropy of the surface diffusion changes alternately
on consecutive terraces.

On the vicinal face, 
two types of step instabilities, 
step wandering and step bunching, occur
when  a specimen is heated by direct 
electric current~\cite{livin-kl91,Latyshev-la98ass}.
The step wandering occurs with step-up current in the region
of relatively large inclination (the tilting angle is
$0.08^\circ \le \theta  \le 0.5^\circ$)~\cite{Nielsen-pp01ss}.
Due to the step wandering, grooves perpendicular to the steps 
appear on the vicinal face.
The step bunching occurs irrespective of the current direction
in the region of small  inclination ($\theta \le 0.08^\circ$).
The  type of dominant terraces, which separate step bunches,
is $\mathrm{T_B}$ with step-down current
and $\mathrm{T_A}$ with step-up current.
The size of bunch increases with time as $t^{1/2}$~\cite{Latyshev-la98ass},
which is independent of the drift direction.
The growth rate of the bunches with  step-down current seems slightly 
slower than that with step-up current.

The step instabilities are caused by drift of adatoms induced 
by the current.
By taking account of alternation of the anisotropic surface diffusion,
the step instabilities are theoretically explained.
If the repulsive interaction is strong so that the step bunching is 
suppressed, the step wandering occurs with 
step-up drift~\cite{Sato-ushprb03},
and straight grooves parallel to the drift appears by  in-phase wandering.
The motion of the steps is given by the solution of
the nonlinear equation derived by Pierre-Louis and 
co-workers~\cite{Pierre-Louis-mskp98prl}.
The step bunching occurs irrespective of the direction of the
drift~\cite{Stoyanov-90jjap,Natori-fy92jjap,Natori-ff92ass,Sato-us02jcg,%
Sato-umh03jpsj,Sato-muh04jpsj}.
Since the current and the drift are
 in the same direction~\cite{Ichikawa-doi92apl,Metois-hp99ss},
the results are consistent with the 
experiments~\cite{livin-kl91,Latyshev-la98ass,Nielsen-pp01ss}.
In a one-dimensional step flow model~\cite{Sato-muh04jpsj},
the size of bunches increases with time as $t^{\beta}$ 
with $\beta$ slightly smaller than $1/2$,
consistent with the experiment~\cite{Latyshev-la98ass}.
However, the growth rate with step-up drift is slower than that 
with step-down drift since terraces with fast diffusion in $y$-direction
are dominant with step-down drift.
This is in contradiction with the experiment~\cite{Latyshev-la98ass}.
Since the model with alternating diffusion anisotropy has explained
the bunching and the wandering instabilities on the Si(001) vicinal
face consistently, 
this disagreement is a major obstacle to the unified understanding.

In the one-dimensional step flow 
models~\cite{Sato-umh03jpsj,Sato-muh04jpsj},
the motion of step bunches with step-down drift is similar to 
that with step-up drift except the time scale.
In the Monte Carlo simulation, however, the step pattern 
is changed by the drift direction~\cite{Sato-us02jcg,Sato-ush03prb}:
the step bunches with step-up drift wander 
more than that with step-down drift.
Such a difference  of the step motion in two-dimension
may solve the disagreement in the growth rate
between the experiment~\cite{Latyshev-la98ass} and the one-dimensional
model~\cite{Sato-umh03jpsj}.
In this paper,
we carry out  Monte Carlo simulations
and show that the difference of the growth rate
vanishes in the two-dimensional model.

\section{Model}\label{sec:model}
For simplicity, we use a square lattice model
with the lattice constant $a=1$.
We take $x$-axis parallel to the steps
and  $y$-axis in the down-hill direction.
Boundary conditions are periodic in the $x$-direction
and helical in the $y$-direction.
We forbid two-dimensional nucleation and use solid-on-solid steps,
i.e. the step positions are single valued functions of $x$.

We repetitively select a solid atom at the step or an adatom 
on the terrace. 
We perform the diffusion and solidification trial for the adatom
and melting trial for solid atom.
In the diffusion trial,
the  adatom hops to a neighboring site.
The anisotropy of the diffusion coefficient and the drift
of adatoms are taken into account in the hopping probability.
On $\mathrm{T_A}$, where the surface diffusion 
in the $x$-direction is faster,
an adatom on the site $(i,j)$ moves to $(i  \pm 1,j)$ with
the probability $1/4$
and to $(i , j \pm 1)$ with the probability 
$p_\mathrm{d} (1 \pm  Fa /k_\mathrm{B}T)/4$,
where $p_\mathrm{d}(<1)$ is the ratio of the two diffusion coefficients.
$F$ is the force to cause the drift.
$F>0$  represents the drift in the down-hill direction.
On $\mathrm{T_B}$, where the surface diffusion in the $y$-direction is faster, 
an adatom on the site $(i,j)$ moves to $(i \pm 1, j)$
with the probability $p_\mathrm{d} /4$ 
and to $(i,j\pm 1)$ with the probability $ (1 \pm  Fa /k_\mathrm{B}T)/4$.
For a diffusion trial,
the time increment is $\Delta t = 1/4N_\mathrm{a}$,
where $N_\mathrm{a}$ is the number of adatoms
so that the fast diffusion coefficient is unity.

If the adatom comes in contact with a step from the lower terrace 
after a diffusion trial,
solidification occurs with  the probability
	\begin{equation}
		p_\mathrm{s}
		=
		\left[
		1
		+
		\exp \left( \frac{\Delta E_\mathrm{s} -\phi }{k_\mathrm{B}T}
		\right)
		\right]^{-1}.
	\end{equation}
$\Delta E_\mathrm{s}$ is given 
by $\Delta E_\mathrm{s} = \epsilon \times$(the increment
of the step perimeter)
and $\phi$ is decrement of the chemical potential by solidification.
The step stiffness $\tilde{\beta} $ is related to $\epsilon$ as
	\begin{equation}
		\frac{2\tilde{\beta}}{k_\mathrm{B}T} 
		= \sinh^2 \frac{\epsilon}{k_\mathrm{B}T}.
	\end{equation}

If we select a solid atom, melting trial is performed.
When an adatom is absent on the top of the  solid atom,
melting occurs with the probability
	\begin{equation}
		p_\mathrm{m}
		=
		\left[
		1
		+
		\exp \left( \frac{\Delta E_\mathrm{s} +\phi }{k_\mathrm{B}T}
		\right)
		\right]^{-1}.
	\end{equation}
There is no extra diffusion barrier over the steps:
steps are permeable.

\section{Results of simulation}\label{sec:results}

In our simulations,
we neglect the long-range repulsive interaction 
between steps,
but take into account a short-range repulsive interaction
by forbidding overlap of steps.
Impingement of atoms and evaporation are absent.

\begin{figure}[thbp]
       \centerline{
       \includegraphics[width=0.5\linewidth]{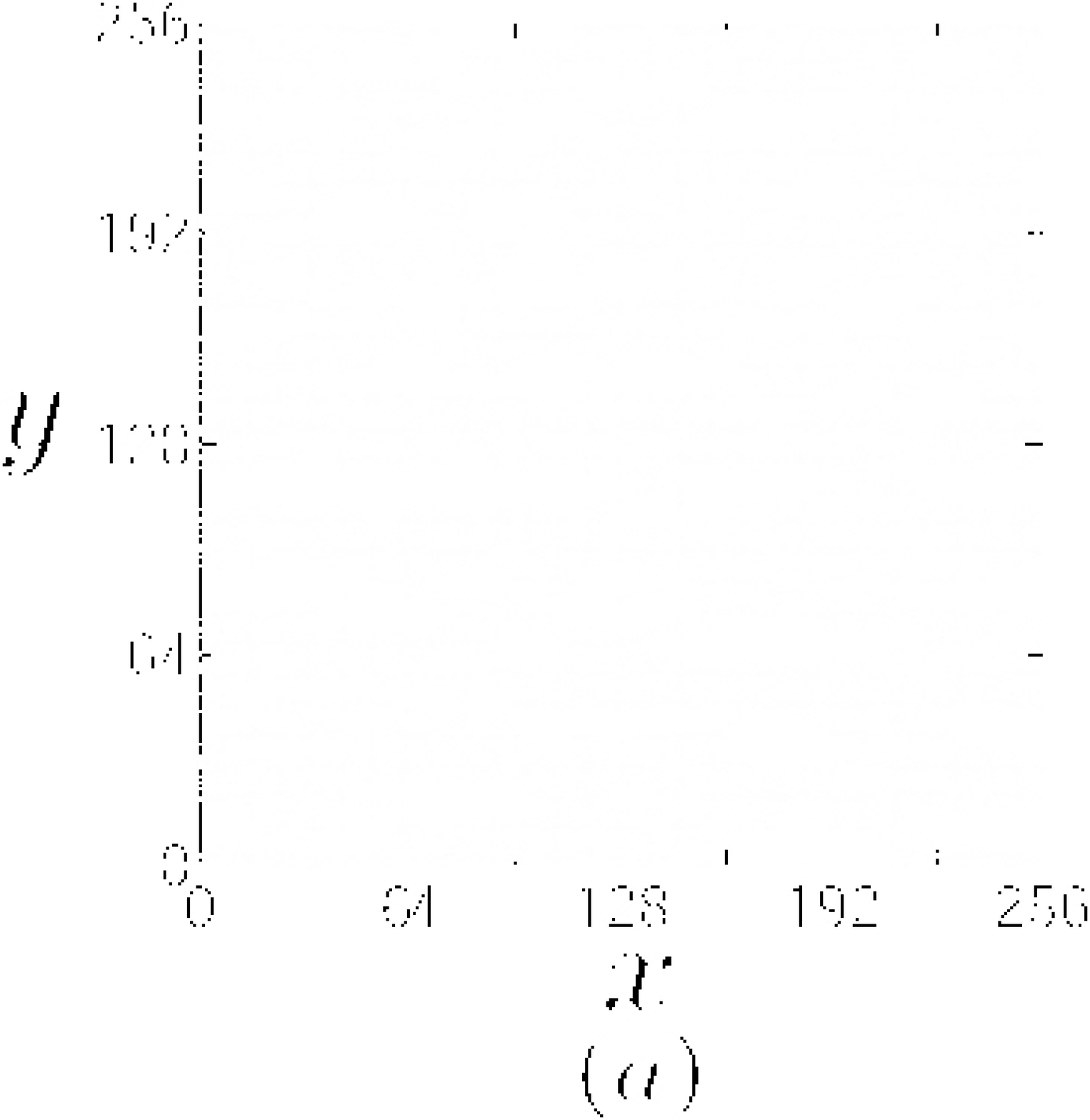}
       \includegraphics[width=0.5\linewidth]{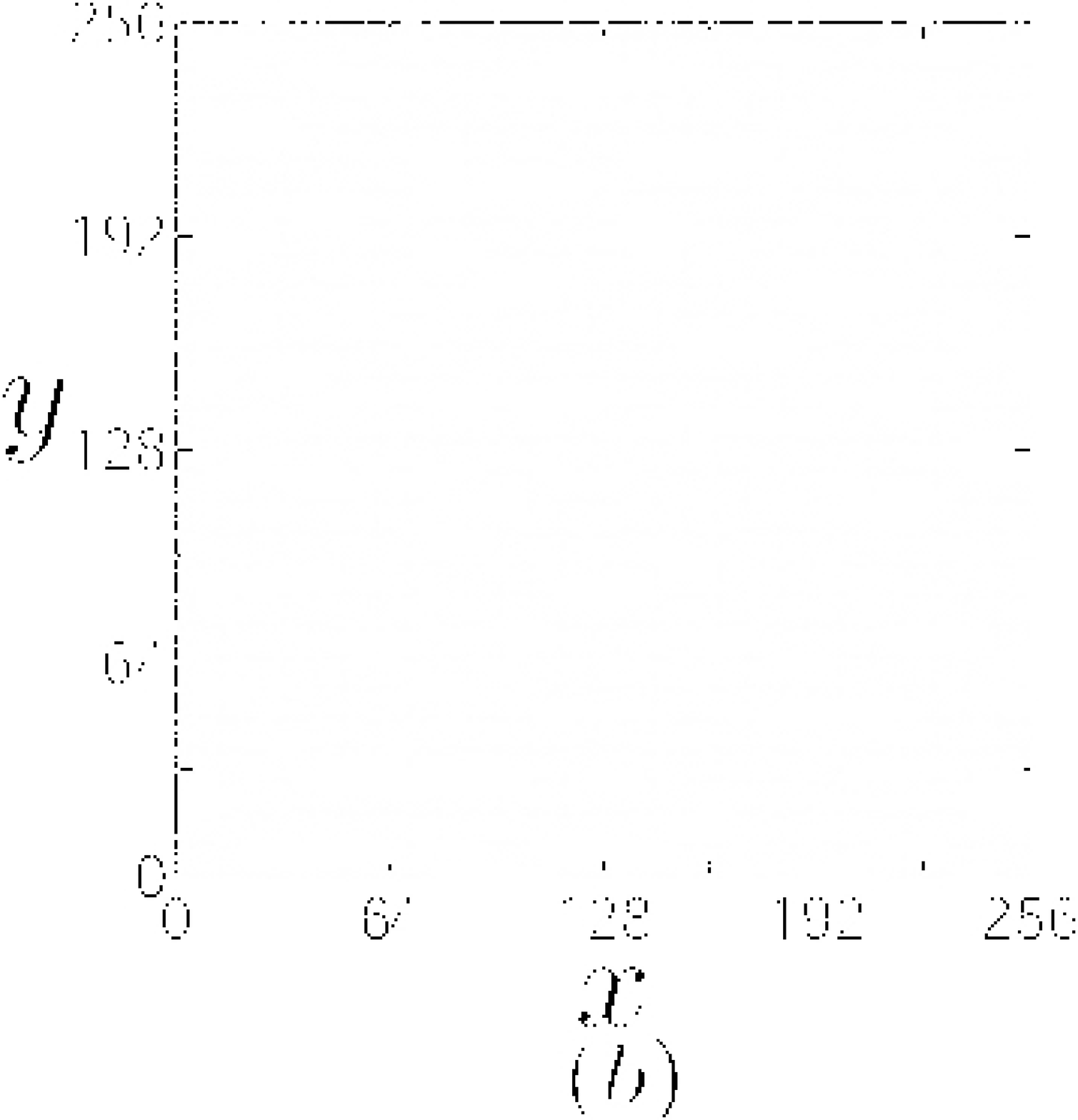}
				}
       \caption{
       Snapshots of step bunching 
       at $t=7.1 \times 10^2$ 
       (a) with step-down drift
       and 
       (b) with step-up drift
       }\label{fig:smallsizeinit}
       \end{figure}

\begin{figure}[thbp]
       \centerline{
       \includegraphics[width=0.5\linewidth]{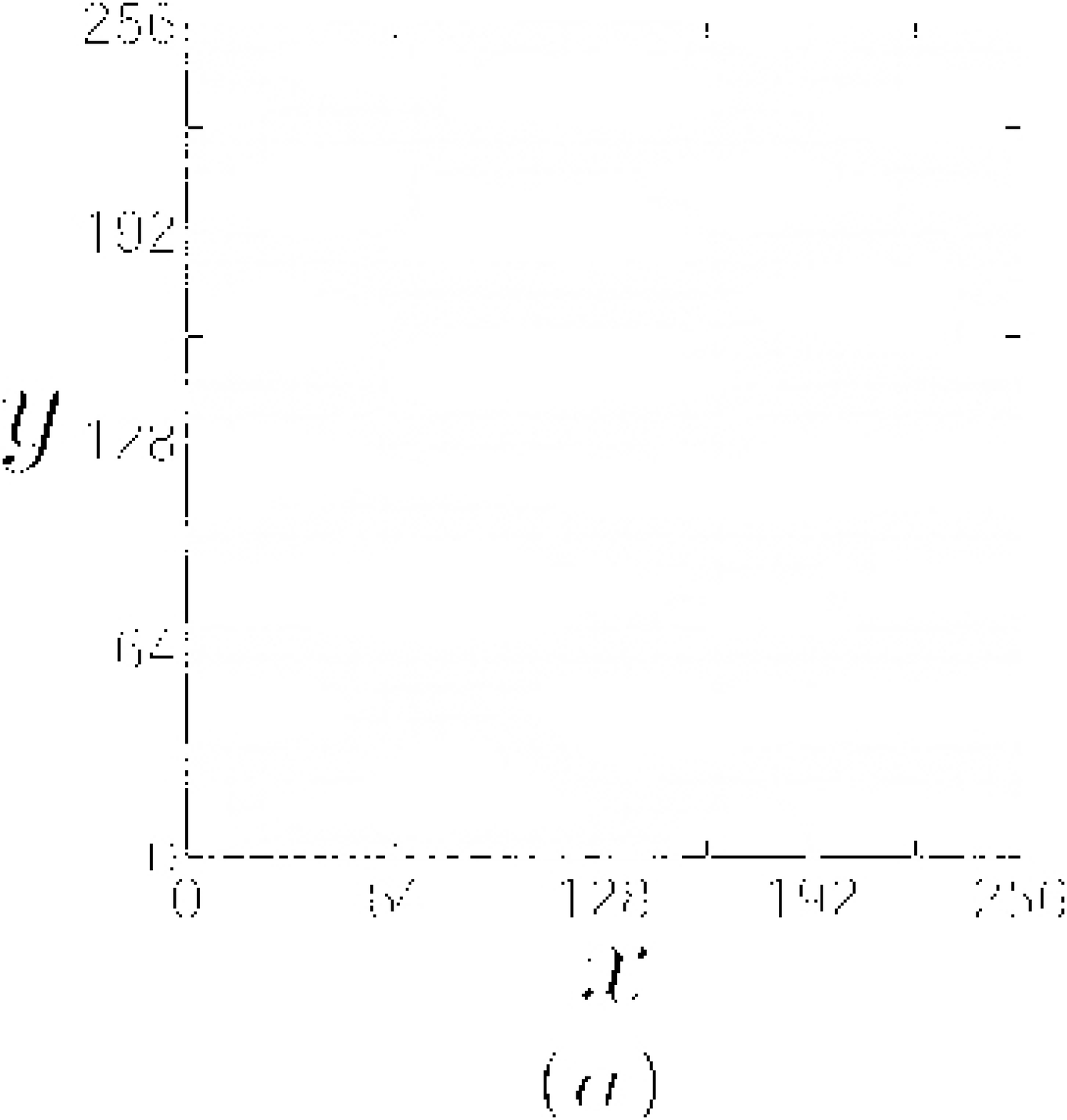}
       \includegraphics[width=0.5\linewidth]{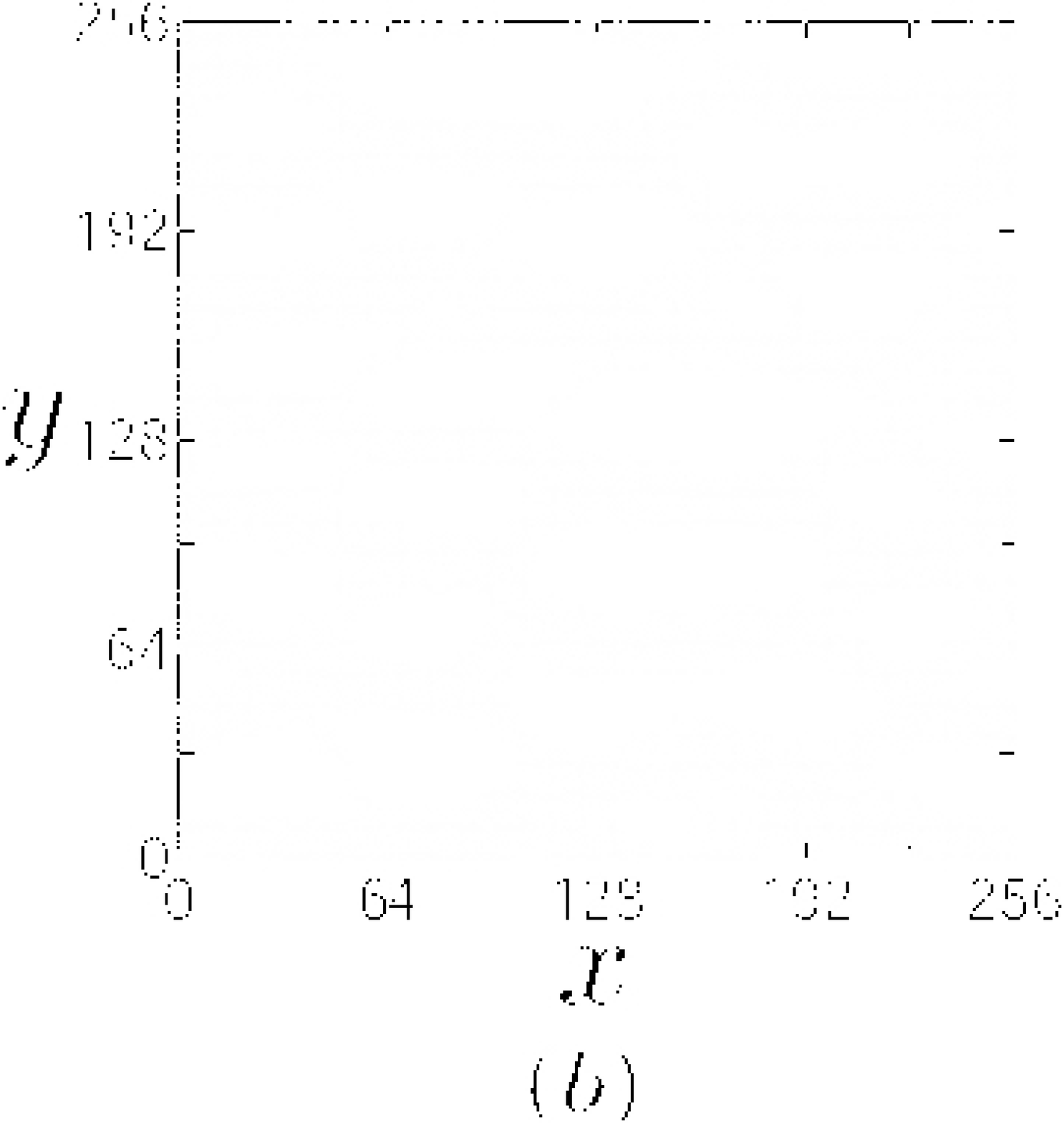}
				}
       \caption{
       Snapshots of step bunching
       (a) with step-down drift at $t= 3.6 \times 10^4$
       and 
       (b) with step-up drift at $t= 3.5 \times 10^4$.
       }\label{fig:smallsize}
       \end{figure}

Figures~\ref{fig:smallsizeinit} and \ref{fig:smallsize} represent
snapshots of the step bunching.
System size is $256 \times 256$
and the number of steps is $64$.
The parameters are 
$\epsilon/k_\mathrm{B}T =1.0$, $\phi/k_\mathrm{B}T =1.5$,
$Fa/k_\mathrm{B}T=  \pm 0.08$ and $p_\mathrm{d}=0.5$.
Initially a few adatoms are present on the vicinal face.
The dotted  lines represent S$_\mathrm{A}$ steps
and the solid lines represent S$_\mathrm{B}$ steps.

The vicinal face is unstable with the drift of adatoms.
Pairing of $\mathrm{S_A}$ and $\mathrm{S_B}$ occurs
in the initial stage.
The upper side step in a pair is $\mathrm{S_A}$
with step-up drift
and $\mathrm{S_B}$ with step-down drift.
Small bunches are formed by coalescence of step pairs.
Since the stiffness is small,
the bunches wander and connect with each other
at many places (Fig.~\ref{fig:smallsizeinit}).
From the figures
one may have impression that he step with step-down drift are more straight.

%
The effect of the drift direction on
the form of bunches becomes evident 
in a late stage (Fig.~\ref{fig:smallsize}).
With step-down drift, the  bunches are 
straight and there are  few  recombination of bunches.
With step-up drift, 
the wandering width of the bunches is large.
The bunches collide with each other and 
frequent recombination is seen.
The difference of the form may affect the time evolution
of bunch size.

\begin{figure}[thbp]
       \centerline{
       \includegraphics[width=0.62\linewidth]{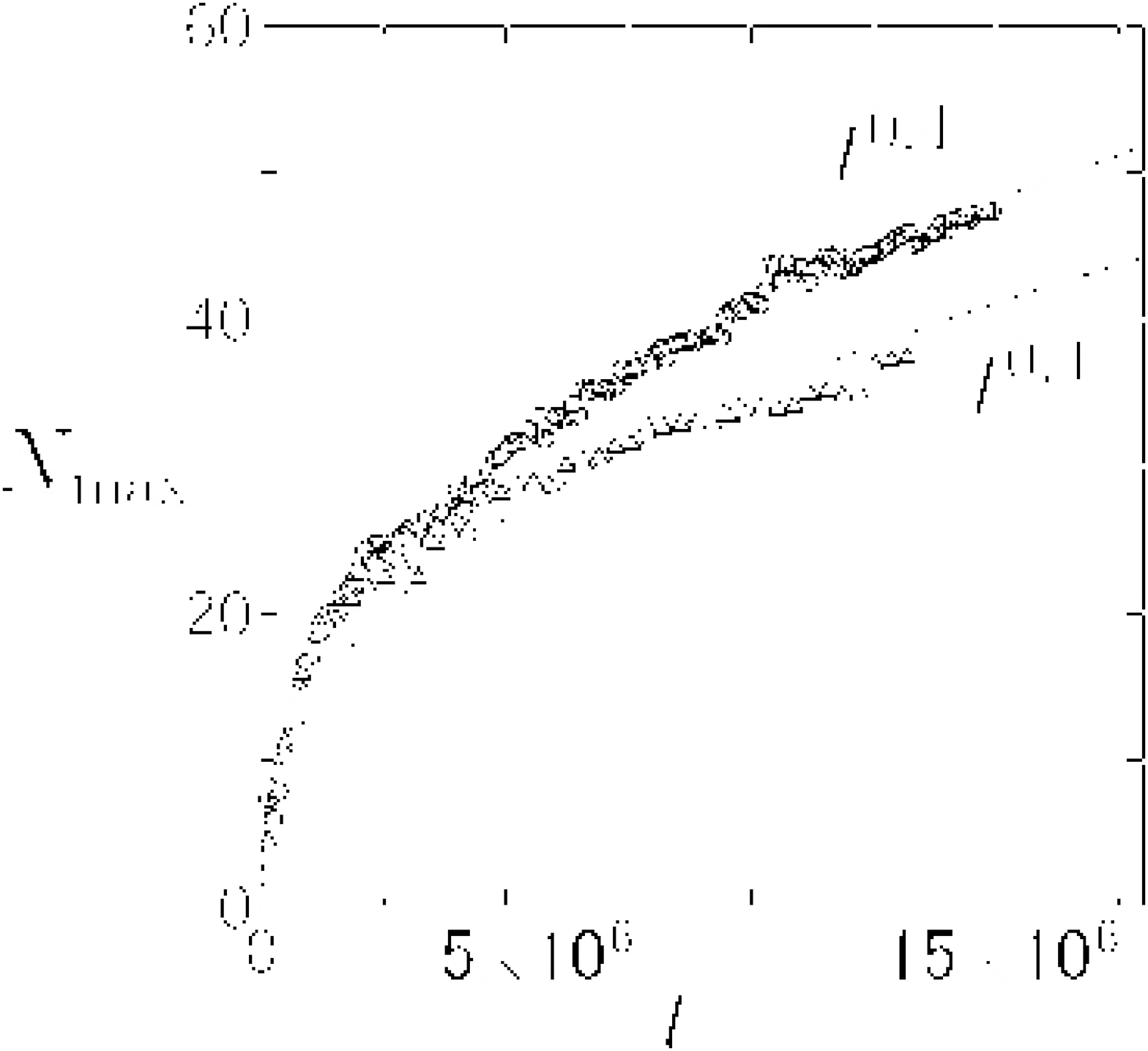}
				}
       \caption{
       Time evolution of the largest bunch size
       with step-up drift ($\bigtriangleup$)
       and  step-down drift ({\Large $\circ$}).
	   The system size is $16 \times 512$
       and the number of steps is $128$. 
       The result is obtained by averaging  over 20 runs.
       }\label{fig:widthsmall}
       \end{figure}

To test the effect of the wandering and recombination on the growth rate, 
we carry out simulations  with a narrow  
system (Fig.~\ref{fig:widthsmall}).
The number of steps is $128$ and the system size is $16 \times 512$.
We use the parameter $\epsilon/k_\mathrm{B}T =0.5$
for step bunches to wander easily.
Other parameters  are the same 
as that in Fig.~\ref{fig:smallsize}.
The wandering of step bunches with step-up drift is suppressed
because of the narrow system width.
The step bunches are straight irrespective of the drift direction.
%

In Figure~\ref{fig:widthsmall},
the number $N_\mathrm{max}$ of steps in the largest bunch at 
$x=1$ is plotted as a function of time.
The bunches grow by the collisions of straight bunches
due  to the fluctuation of position 
of bunches~\cite{Sato-umh03jpsj,Sato-muh04jpsj}.
The growth rate with step-up drift is 
slower than that with step-down drift 
like the one-dimensional model.
$N_\mathrm{max}$ seems to increase in the power law,
$N_\mathrm{max} \sim t^{\beta}$ with $\beta \approx 0.4$~\cite{criterion},
which is consistent with the one-dimensional result with a 
combination of $\ln r $ and $r^{-2}$
potentials~\cite{Sato-muh04jpsj}.

\begin{figure}[thbp]
       \centerline{
       \includegraphics[width=0.64\linewidth]{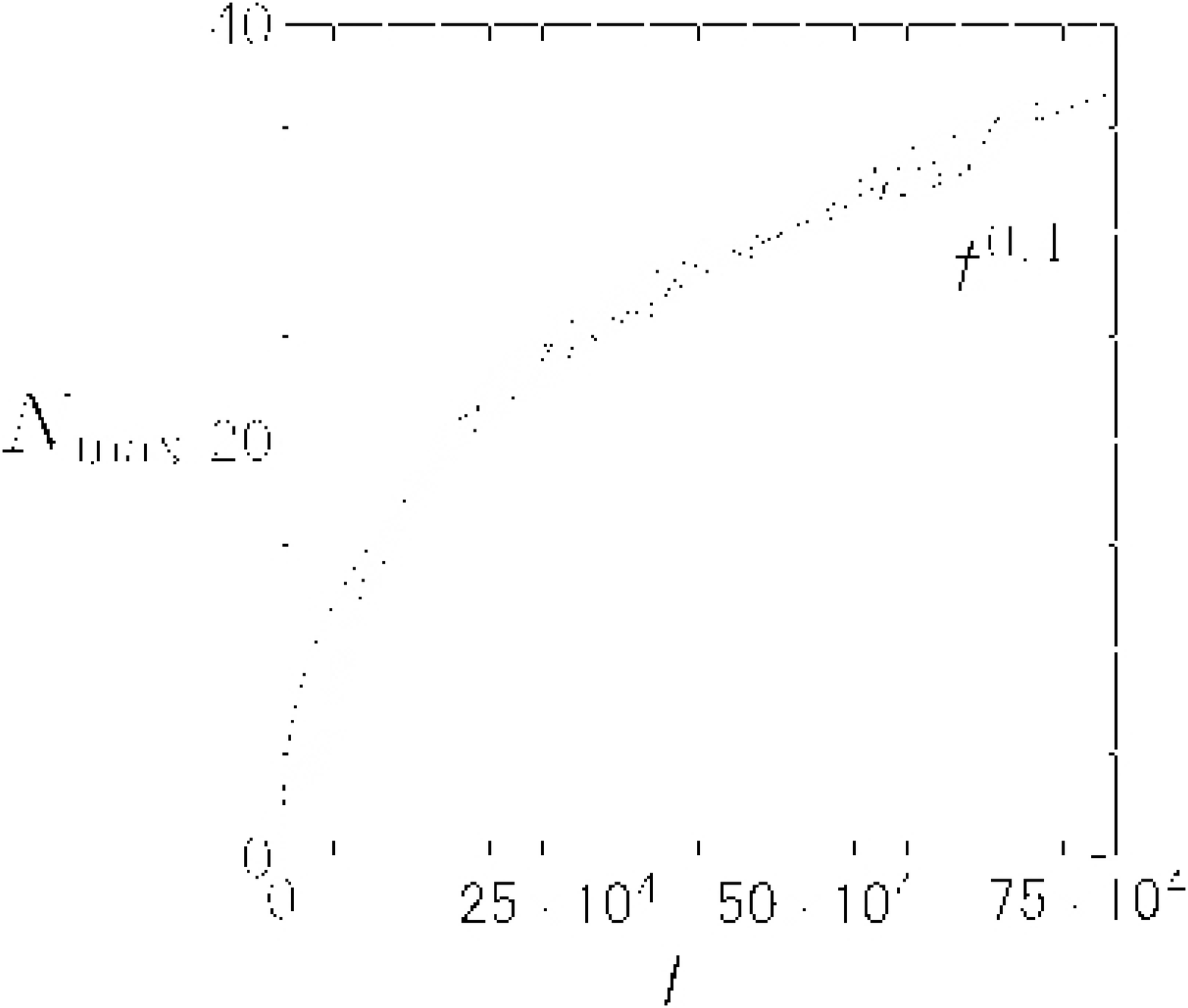}
				}
       \caption{
       Time evolution of the largest bunch size
       with step-up drift ($\bigtriangleup$)
       and  step-down drift ({\large $\circ$}).
	   The system size is $512 \times 512$
       and the number of steps is $128$. 
       The result is obtained by averaging  over 20 runs.
       }\label{fig:width}
       \end{figure}

We carry out simulations with a large
system size: $512 \times 512$ (Fig.~\ref{fig:width}).
In contrast to the  one-dimensional model~\cite{Sato-muh04jpsj},
the growth of bunch size with step-up drift is as fast as
that with step-down  drift. 
Thus the slow diffusion in $y$-direction is compensated by
the efficient coalescence by the wandering of bunches.

%

In the experiment~\cite{Latyshev-la98ass}, 
however, the step bunching with step-up current  seems 
slightly faster than that with step-down current.
Additional mechanism to accelerate the bunching with step-up drift
may be required.
On the Si(001) vicinal face,
$\mathrm{S_B}$ steps are rougher than $\mathrm{S_A}$ steps
and the kinetic coefficient is probably larger for $\mathrm{S_B}$.
We  take account of  the difference as the  probabilities  
of solidification and melting.
At $\mathrm{S_A}$,
the probability of solidification is assumed to be
$r p_\mathrm{s}$ and that of melting is $r p_\mathrm{m}$
with $r \le 1$.

\begin{figure}[thbp]
       \centerline{
       \includegraphics[width=0.64\linewidth]{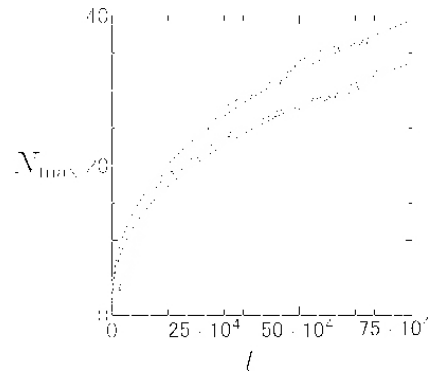}
				}
       \caption{
       Time evolution of the largest bunch size
       with step-up drift ($\bigtriangleup$)
       and  step-down drift ({\large $\circ$}).
       The step kinetics of $\mathrm{S_A}$ is slower ($r =0.1$)
	   The system size is $512 \times 512$
       and the number of steps is $128$. 
		The result is obtained by averaging  over 30 runs.
       }\label{fig:kineticswidth}
       \end{figure}

Figure~\ref{fig:kineticswidth}
represents the time evolution of bunch size
with $r=0.1$.
Other parameters and system size are the same
as those in Fig.~\ref{fig:width}.
The growth exponent does not  change significantly.
The growth rate with step-down drift is suppressed as expected
from the reduction factor $r$,
but that with step-up drift is slightly enhanced.

\begin{figure}[thbp]
       \centerline{
       \includegraphics[width=0.50\linewidth]{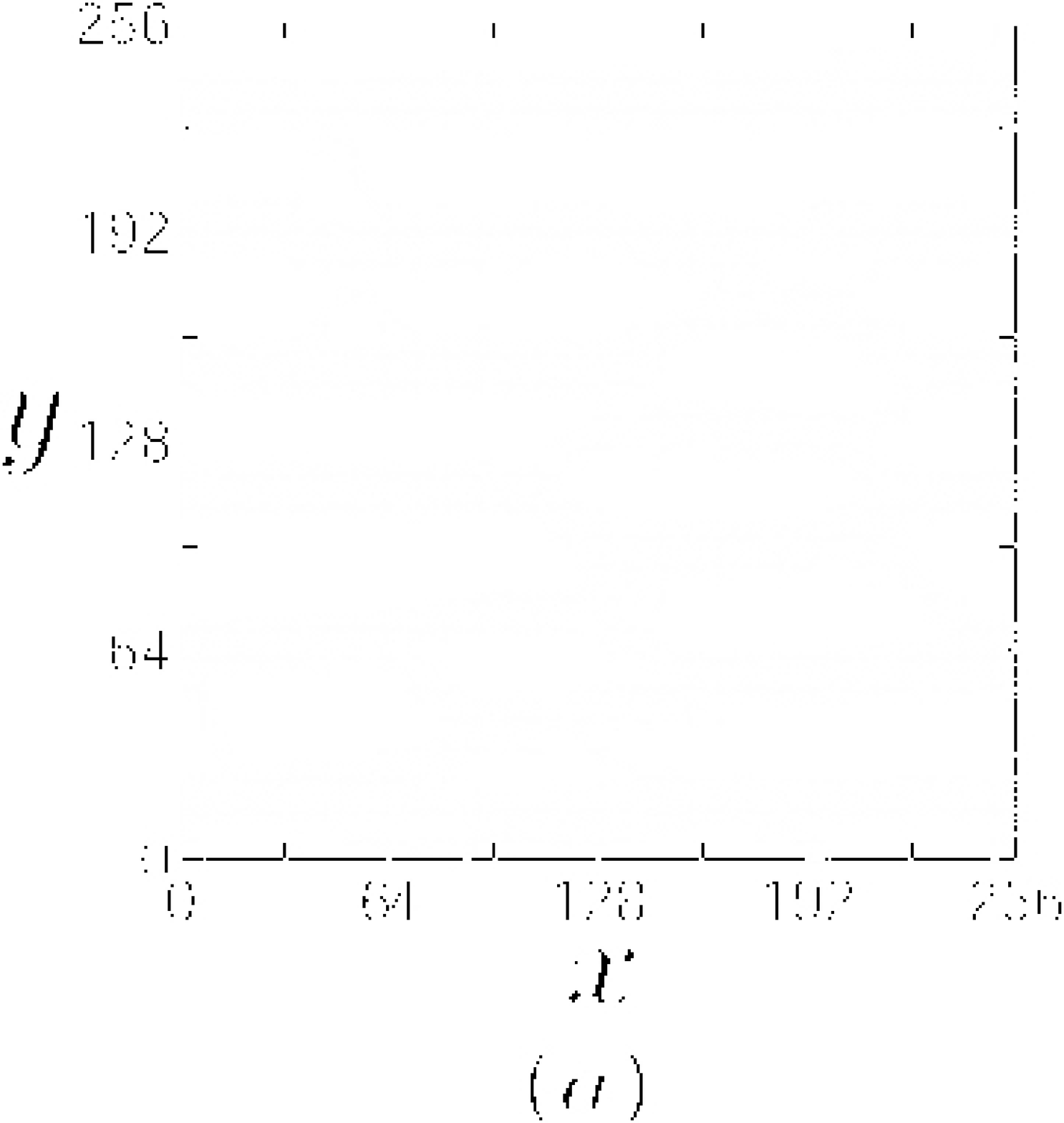}
       \includegraphics[width=0.50\linewidth]{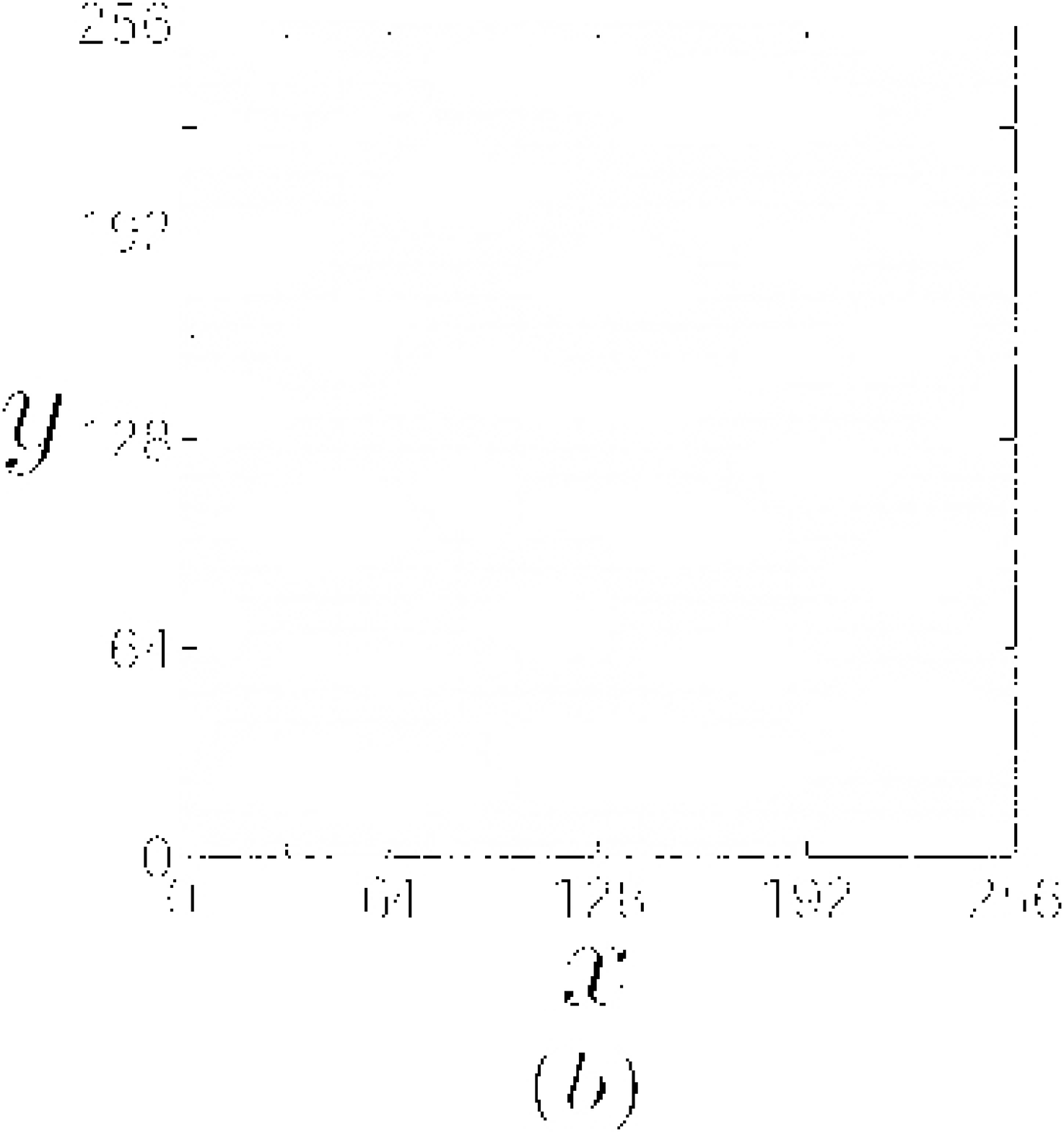}
				}
       \caption{
       Snapshots of step bunching
       (a) with step-down drift at $t= 3.6 \times 10^4$
       and 
       (b) with step-up drift at $t= 3.5 \times 10^4$.
       }\label{fig:r-snapshot}
       \end{figure}

In figure~\ref{fig:r-snapshot} we show that the step patterns.
As explained already, the bunches with step-down drift
are straight.
The bunches with step-up drift are wavy
and shows more recombination patterns.
The reduction of kinetic coefficient of $\mathrm{S_A}$ seems
to enhance the waviness. 
{ As a result, 
the growth rate with step-up drift becomes} faster than
that with step-down drift, 
which is in agreement with  the experiment~\cite{Latyshev-la98ass}.

\section{Summary and Discussion}\label{sec:summary}
We studied the time evolution of bunch size by Monte Carlo simulation.
With step-down drift, 
the step bunches are  straight and 
there are  few  recombination of  bunches.
With step-up drift, the bunches wander
and recombinations  of bunches occur frequently

In the one-dimensional model~\cite{Sato-umh03jpsj,Sato-muh04jpsj},
the  velocity of a bunch is roughly proportional
to the diffusion coefficient of $y$-direction in large terraces.
Bunches  with step-down drift grow faster
than those  with step-up drift,
which is in disagreement with the experiment~\cite{Latyshev-la98ass}.
In the two-dimensional model,
bunches with step-up drift wander and 
collide with each other more frequently
than those with step-down drift.
Since coalescence of bunches starts from the connected parts,
the step bunching with step-up drift occurs
more frequently than  that with step-down drift.
Also the fast transverse diffusion with step-up drift helps
recombination of bunches.
Consequently the growth rate of bunches  with step-up drift is as
fast as that with step-down drift.

When the difference of the kinetic coefficient is taken into account,
the bunches with step-up drift grows faster
than those with step-down drift.
The reversal of the growth rate is thus attributed to the two-dimensional
step pattern.
Since we have not succeeded in quantifying the effect,
it is not clear if the proposed effect  is sufficient to explain 
the experiment~\cite{Latyshev-la98ass}.

\acknowledgments
This work was supported by Grant-in-Aid for Scientific
Research from Japan Society for  the Promotion of Science.
M. U. and Y. S. benefited from the inter-university cooperative
research program of the Institute for Materials Research, Tohoku University.



\begin{thebibliography}{999}
\bibitem{criterion}
In the limited size of our simulation,
it is difficult to define a bunch size.
We regard a step pair as being contained in the bunch
if the distance between the step pair and the  bunch
is smaller than a prescribed value.
The obtained exponent,
though not very sensitively,
depends on the value.
\end{thebibliography}
\end{document}